# MoSi$_2$N$_4$: An emerging 2D electronic material with protected band edge states and dielectric tunable quasiparticle and optical properties

Yabei Wu,[1,2,3,‡] Zhao Tang,[4,‡] Weiyi Xia,[4] Weiwei Gao,[5] Fanhao Jia,[4,6] Yubo Zhang,[1,2] Wenguang Zhu,[3] Wenqing Zhang,[1,2,†] and Peihong Zhang[4,†]

1. Department of Physics and Shenzhen Institute for Quantum Science & engineering, Southern University of Science and Technology, Shenzhen, Guangdong 518055, China

2. Guangdong Provincial Key Lab for Computational Science and Materials Design, and Shenzhen Municipal Key Lab for Advanced Quantum Materials and Devices, Southern University of Science and Technology, Shenzhen, Guangdong 518055, China

3. ICQD, Hefei National Laboratory for Physical Science at the Microscale, Key Laboratory of Strongly-Coupled Quantum Matter Physics, Chinese Academy of Sciences, Department of Physics, and Synergetic Innovation Center of Quantum Information and Quantum Physics, University of Science and Technology of China, Hefei, Anhui 230026, China

4. Department of Physics, University at Buffalo, State University of New York, Buffalo, New York 14260, USA

5. Department of Physics, Dalian University of Technology, Dalian, Liaoning 116024, China

6. International Centre for Quantum and Molecular Structures, Materials Genome Institute, Department of Physics, Shanghai University, 99 Shangda Road, Shanghai 200444 China

[†]zhangwq@sustech.edu.cn; pzhang3@buffalo.edu

[‡]These authors contributed equally to this work.

**Abstract**

As it is the case for all low-dimensional materials, the low energy electronic structure of two-dimensional (2D) materials are inherently prone to environmental perturbations. While there are applications (e.g., sensors) that exploit this property, the fact that the band-edge states of 2D materials are susceptible to often unpredictable environmental coupling effects may pose significant challenges to their practical applications in electronic or optoelectronic devices. A 2D material couples with its environment through two mechanisms: local chemical coupling arising from the overlap and hybridization of wave functions, and nonlocal dielectric screening effects that renormalize the Coulomb interaction in the system. The local chemical coupling, which may arise from the (often unintentional) surface adsorption or interfacial chemical coupling, is difficult to predict or control experimentally. Nonlocal dielectric screening effects, on the other hand, can be tuned by choosing the substrates or layer thickness in a controllable manner. Therefore, a compelling 2D electronic material should offer band edge states that are robust against local chemical coupling effects. Here we demonstrate that the recently synthesized $MoSi_2N_4$ is an ideal 2D semiconductor with robust band edge states protected from capricious environmental chemical coupling effects. We have carried out thorough density functional theory (DFT) and many-body perturbation theory (MBPT) calculations to illustrate how the band edge states of $MoSi_2N_4$ are essentially shielded from direct chemical coupling to its environment, and, at the same time, the quasiparticle and optical properties of $MoSi_2N_4$ can be modulated through the nonlocal dielectric screening effects. This unique property, together with the moderate band gap and the thermodynamic and mechanical stability of this material, may pave the way for a range of applications in areas including energy, 2D electronics, and optoelectronics.

# I. Introduction

It is difficult to overstate the research interest in two-dimensional (2D) materials due to their rich physics and potential applications in next-generation electronic devices [1, 2]. While significant progress has been made in our understanding of the fundamental physics and properties of 2D materials, their practical applications have certainly lagged behind. Much recent effort has been put into exploiting the interlayer coupling effects of 2D materials to tune their electronic, optical, or magnetic properties [3-6], or even to create exotic states such as those observed in twisted graphene [7-9].

However, the fact that the low energy electronic structure, especially the band edge states, of 2D materials are prone to environmental perturbations may become a major issue facing their practical applications in electronic or optoelectronic devices. There are two fundamental mechanisms through which a 2D material may couple with its environment: local chemical interaction and nonlocal dielectric screening. Nonlocal screening effects can be engineered with the choice of the substrate and/or the layer thickness, which can actually serve as an advantageous mechanism to tailor the quasiparticle and optical properties of 2D semiconductors in a controllable manner [10-12]. Local chemical coupling, which may arise from unintentional surface adsorption and/or interfacial chemical coupling, on the other hand, is particularly difficult to predict and control in experiment but can significantly affect the band edge states of 2D materials. Note that even very weak chemical coupling at typical van der Waals (vdW) distances, i.e., without the formation of conventional chemical bonds, can significantly affect the band edge states. These interactions may strongly modify the electronic structure of 2D semiconductors, thus their device performance. Therefore, a compelling 2D electronic material should offer band edge states that are robust against weak surface or interfacial chemical interactions.

Here we demonstrate that the recently synthesized $MoSi_2N_4$ [13] may provide such an ideal 2D semiconductor with band edge states being protected from capricious chemical couplings. $MoSi_2N_4$ belongs to a family of emerging 2D materials with the chemical formula $MSi_2N_4$, where M is a transition metal and Si and N may be replaced by other group IV and V elements, respectively. This material has demonstrated excellent ambient stability [13]. A number of interesting properties such as piezoelectricity and high thermal conductivity [14], and spin-valley coupling [15, 16] have also been predicted. In this work, we have carried out thorough density functional theory (DFT) and many-body perturbation theory calculations to illustrate how the band edge states of $MoSi_2N_4$ are essentially shielded from direct chemical coupling to its environment, and at the same time, the quasiparticle and optical properties of $MoSi_2N_4$ can be modulated through the nonlocal dielectric screening effects which renormalize the electron-electron (*e-e*) and electron-hole (*e-h*) interactions.

## II. Results and Discussion

### A. Electronic and structural properties of $MoSi_2N_4$: from monolayer to bulk

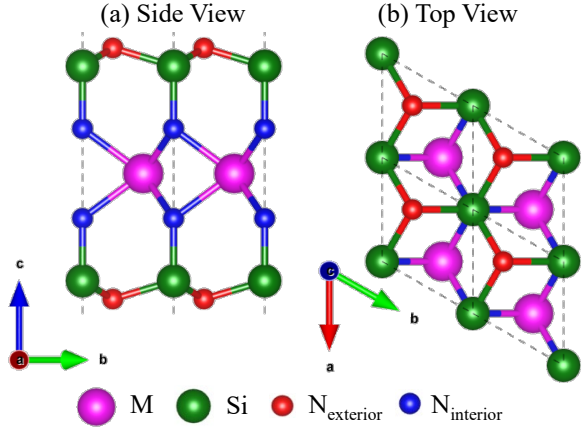

**Figure 1.** Side view and top view of the crystal structure of monolayer $MoSi_2N_4$. The surface and interior N atoms are shown with red and blue colors, respectively.

The crystal structure of the newly synthesized monolayer $MoSi_2N_4$ is shown in Fig. 1, which can be considered as an $MoN_2$ layer sandwiched between two SiN layers. The optimized in-plane lattice constant for monolayer $MoSi_2N_4$ is 2.896 Å, which is consistent with an earlier report [13]. The distance between two Si layers is 5.99 Å, which also agrees with experiment [13]. The strong polar covalent bonding and the relatively thick layer structure provide an energetically and mechanically stable structure for potential applications.

Such a multi-atomic layer structure also offers the possibility of having protected band edge states from weak interfacial chemical coupling if these states are primarily derived from orbitals of the interior atoms. To investigate such possibility, we carry out DFT calculations of the layer-dependent electronic structure of $MoSi_2N_4$. We first construct six bilayer structures and the corresponding bulk structures with different stacking patterns. The structures are optimized using the PBE functional with the DFT-D3 [17] correction to account for the interlayer vdW interaction. Details of the optimized structures are shown in the Supplementary Material in Fig. S1 and Table S1. The calculated inter-layer separation ranges from 3.270 to 2.825 Å for the bilayer structure, and 3.222 to 2.807 Å for the bulk phase. Available experimental values for the bulk interlayer distances are also shown for comparison. Not surprisingly, the lowest energy structures have the smallest interlayer separations. Table I compares the calculated interlayer separation $d$ and binding energy of the most stable bilayer and bulk $MoSi_2N_4$ with a few selected layered materials. Compared with other layered materials, $MoSi_2N_4$ has slightly smaller interlayer distances $d$, and the interlayer binding energy also falls within that of a typical layered material.

**Table I.** Comparison between the calculated interlayer binding energy (in meV/Å$^2$) and interlayer separation $d$ (in Å) of $MoSi_2N_4$ with a few selected layered materials using the PBE functional with the DFT-D3 correction.

|  | $MoSi_2N_4$ | $MoS_2$ | $WS_2$ | Black phosphorus | $C_3N$ | $C_3B$ |
|---|---|---|---|---|---|---|
| Binding Energy | 35.7 | 32.5 | 40.0 | 32.3 | 29.7 | 27.2 |
| Bilayer $d$ | 2.825 | 2.944 | 3.012 | 3.182 | 3.202 | 3.206 |
| Bulk $d$ | 2.807 | 2.921 | 2.951 | 3.194 | 3.187 | 3.097 |
|  | -- | 2.975 [18] | 3.019 [19] | 3.101 [20] | -- | -- |

It is well known that interlayer chemical coupling can significantly modify the low energy electronic structure, especially the band gap, of many vdW layered materials. Such chemical coupling effects can be well described within DFT provided that the interlayer separations are properly determined. With the small interlayer separations and moderate interlayer binding energy, we would expect that $MoSi_2N_4$ experience similar chemical coupling effects found in other layered materials.

Figure 2 compares the PBE band structure of monolayer, bilayer, and bulk $MoSi_2N_4$ calculated using the optimized structures. Monolayer $MoSi_2N_4$ has an indirect band gap of 1.79 eV calculated within PBE; the valence band maximum (VBM) locates at the Γ point and the conduction band minimum (CBM) locates at K. These results are consistent with other theoretical results [13, 15]. Surprisingly, the calculated PBE band gap of bilayer MoSi$_2$N$_4$ is 1.77 eV, and that of the bulk phase is 1.70 eV, which are only 20 meV and 90 meV smaller than that of the monolayer structure, respectively.

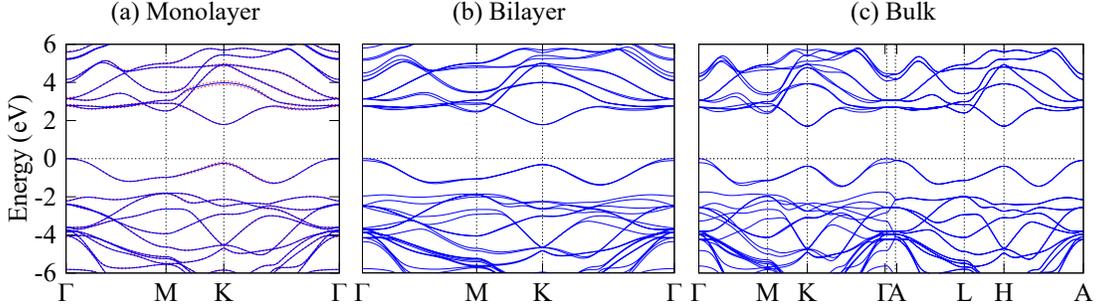

**Figure 2.** DFT-PBE calculated band structures for MoSi$_2$N$_4$ from monolayer to bulk. The blue solid and red dotted lines in (a) represent the band structure without and with SOC effects for monolayer MoSi$_2$N$_4$, respectively.

The dispersion of the low energy electronic structures also shows negligible changes going from monolayer to bulk. The fact that interlayer coupling has negligible effects on the calculated band gap and the low energy band dispersion (at the DFT level) of MoSi$_2$N$_4$ is in stark contrast with other layered materials, as shown in Table II, considering that these materials have comparable interlayer separations and binding energies (Table I). For example, the DFT band gap of MoS$_2$ changes from 1.72 eV (monolayer) to 0.88 eV (bulk), and C$_3$N and C$_3$B both show a band gap change of over 1.3 eV from monolayer to bulk.

**Table II.** Interlayer chemical coupling effects on the band gap of MoSi$_2$N$_4$ compared with other layered materials. The band gaps (in eV) are calculated at the DFT-PBE level.

|  | MoSi$_2$N$_4$ | MoS$_2$ [21] | WS$_2$ [21] | Black phosphorus [5] | C$_3$N [22] | C$_3$B [3] |
|---|---|---|---|---|---|---|
| Monolayer | 1.79 | 1.72 | 1.96 | 0.83 | 0.39 | 0.64 |
| Bilayer | 1.77 | 1.26 | 1.45 | 0.48 | 0.13 | 0.10 |
| Bulk | 1.70 | 0.88 | 0.98 | 0.04 | -0.98 | -0.68 |

Before we proceed, we briefly discuss the spin-orbit coupling (SOC) effect in this system. The PBE band structure of monolayer MoSi$_2$N$_4$ calculated with SOC included is shown with red dotted lines in Fig. 2(a). The SOC effects result in a splitting of 130 meV of the top valence band at the K point; the SOC effects on the CBM (at the K point) and VBM ($\Gamma$) states are negligible. These results agree well with experiment [13].

## B. Robust band edge states protected from interfacial chemical coupling

The negligible layer-dependence of the calculated DFT gap strongly hints at the presence of robust near-edge electronic states that are shielded from interfacial chemical coupling. This is only possible if the band edge states are derived primarily from interior (i.e., Mo and interior N atoms) atomic orbitals. In order to gain further understanding, we show in Fig. 3 (a) the decomposition of the Kohn-Sham wave functions near the band gap into contributions from atomic orbitals. It is obvious that the band edge states are largely derived from Mo $d$ orbitals with small admixture of other atomic orbitals. This is also clearly seen in Fig. 3 (b) and (c), in which we show the isosurface charge density plots of the band edge states at the K point: the charge density of the band edges states mainly locates inside the $MoSi_2N_4$ layer. Considering the relatively thick monolayer structure and the fact that Mo $d$ orbitals are fairly localized, interlayer chemical coupling effects on the band edge states are greatly reduced.

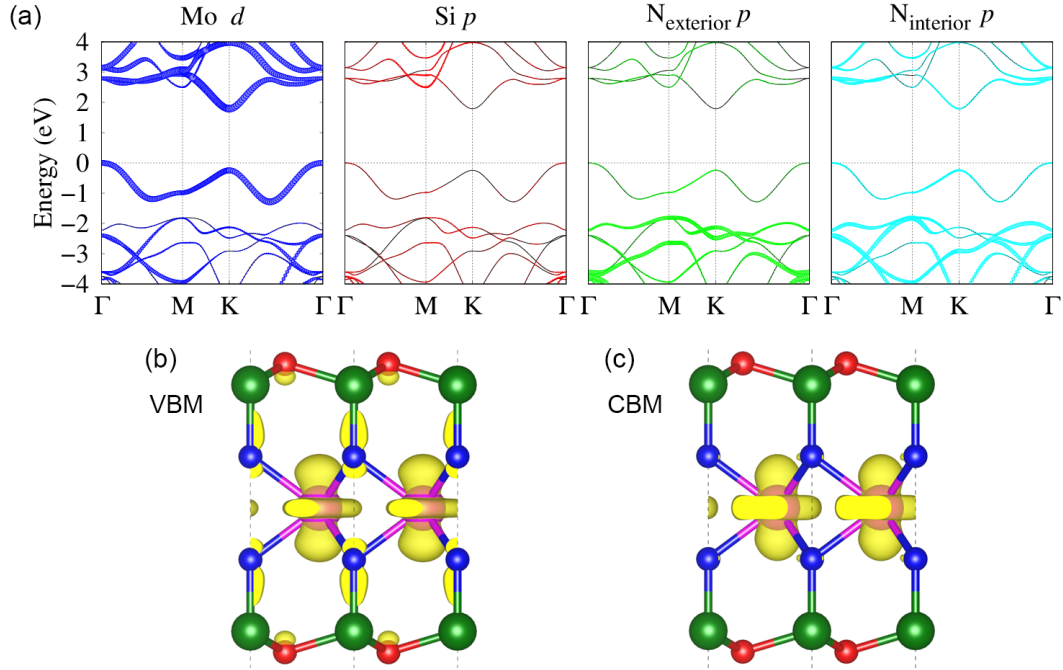

**Figure 3.** Atomic character of the near band edge states of $MoSi_2N_4$. (a) Decomposition of Bloch wave function of $MoSi_2N_4$ into atomic contributions, (b) the isosurface charge density plot ($\rho=0.02$ e/Å$^3$) for the top of the valence band at the K point, and (c) the isosurface charge density plot for the bottom of the conduction band at the K point.

The presence of small admixtures of atomic orbitals derived from Si and surface N atoms in the band edge states explains the slight layer-dependent DFT band structure and band gap as shown in Fig. 2 and Table II. Interlayer chemical coupling (hybridization) will occur for states which have significant contributions from outer atomic orbitals. Interestingly, these states are either significantly above or below the band gap. Therefore, $MoSi_2N_4$ offers an exciting material system with band edge states *protected* from interfacial chemical coupling. It should be pointed out that although we only discuss interlayer chemical coupling effects here, we expect that the band edge states of $MoSi_2N_4$ are similarly protected from other surface or interfacial perturbations such as physical adsorptions, or weak chemical coupling with substrates (i.e., without the formation of strong chemical bonds). As we have mentioned earlier, these perturbations are difficult to control or predict but may significantly affect the performance of 2D electronic devices.

**C. Layer-dependent quasiparticle properties: Tailoring the electronic structure with nonlocal dielectric screening effects**

Our results have unequivocally demonstrated that $MoSi_2N_4$ is a unique layered material with band edge states protected from surface or interfacial perturbations. This does not mean, however, that the quasiparticle or optical properties of $MoSi_2N_4$ are not affected by nonlocal interlayer or substrate coupling effects. In fact, substrate or layer-dependent dielectric screening effects can strongly renormalize the *e-e* or *e-h* interactions, thus the quasiparticle and excitonic properties of 2D materials. Fortunately, unlike local chemical coupling, which is challenging to control in experiment, substrate or interlayer screening effects can serve as an effective means to modulate the quasiparticle and optical properties of 2D materials [10-12]. To this end, we first carry out fully converged GW calculations for monolayer, bilayer, and bulk $MoSi_2N_4$, aiming at illustrating the dielectric screening effects on the quasiparticle properties, in particular, the quasiparticle band gap, of $MoSi_2N_4$.

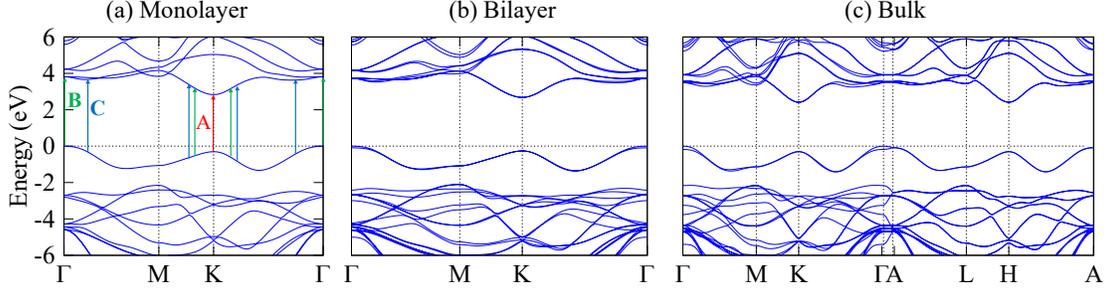

**Figure 4.** GW band structures of monolayer (a), bilayer (b), and bulk (c) MoSi$_2$N$_4$. Important electron and hole states involved in the optical transitions that give rise to the A, B, and C low energy absorption peaks in monolayer MoSi$_2$N$_4$ are also shown.

Figure 4 compares the GW band structures of monolayer, bilayer, and bulk MoSi$_2$N$_4$. The band gaps of all three structures remain indirect ($\Gamma$ to K) in nature within the GW approximation. Although the DFT band gap of MoSi$_2$N$_4$ shows a very small layer dependence, changing from 1.79 to 1.70 eV from the monolayer to the bulk phase as we have discussed earlier, the quasiparticle band gap shows more a substantial layer dependence as summarized in Table III. The calculated GW band gaps (indirect) are 2.82 eV for monolayer, 2.67 eV for bilayer, and 2.41 eV for bulk MoSi$_2$N$_4$, representing a self-energy correction ranging from 1.03 (monolayer) to 0.71 (bulk) eV. Note that even calculated at the GW level, the layer-dependence of the quasiparticle band gap of MoSi$_2$N$_4$ is still significantly weaker than other well-known 2D materials such as MoS$_2$ [23, 24] and black phosphorus [5]. For instance, the GW band gap correction for monolayer MoS$_2$ is 0.96 eV [23], and that for bulk is 0.41 eV [24]. Those for black phosphorus are 1.17 and 0.26 eV [5]. These results again suggest that the band edge states of MoSi$_2$N$_4$ are not as sensitive to environmental perturbation as other 2D materials. On the other hand, controlled fine tuning of the quasiparticle band gap (from 2.82 to 2.41 eV) is still possible with the choice of substrate and/or layer thickness. Similarly, optical properties can be modulated through engineering the dielectric screening, as we will discuss in the next section.

It should be pointed out that fully converged GW calculations for 2D materials remain a challenge due to the large cell size and the unusual analytical behaviors of 2D dielectric functions and electron self-energies, which make conventional GW calculations using the band-by-band summation and the uniform Brillouin zone (BZ) integration approach rather inefficient. Our work takes advantage of the recent code developments [25, 26] that can drastically speed up GW calculations for 2D materials. Using these new developments, we can effectively include *all* conduction bands in the GW calculations, thus eliminating the need for tedious band convergence tests [25]. In addition, the combined subsampling and analytical BZ integration approach [26] greatly improves the efficiency of the BZ integration of the quasiparticle self-energy. In this work, the self-energy integration in the BZ is carried out using a 6×6×1 *k*-grid with four subsampling points; these parameters are sufficient to converge the GW band gap to within 0.03 eV as discussed in our previous publications [3, 22, 26, 27]. We have also tested the convergence of the calculated results with respect to the thickness of vacuum layer. More detail of the convergence test of our GW and BSE calculations can be found in Supplementary Material.

**Table III.** Layer-dependent direct and indirect quasiparticle band gaps (in eV) of MoSi$_2$N$_4$ calculated within the GW approximation.

| **Indirect gap** | Monolayer | Bilayer | Bulk |
|---|---|---|---|
| DFT-PBE | 1.79 | 1.77 | 1.70 |
| GW | 2.82 | 2.67 | 2.41 |
| GW correction | 1.03 | 0.90 | 0.71 |
| **Direct gap (K)** | | | |
| DFT-PBE | 2.03 | 2.09 | 2.10 |
| GW | 3.13 | 2.94 | 2.71 |
| GW correction | 1.10 | 0.85 | 0.61 |

**D. Layer-dependent electron-hole excitations and optical properties**

Now we investigate the layer-dependent *e-h* excitations and optical properties of MoSi$_2$N$_4$. The quasiparticle properties are calculated within the GW approximation as described in the previous section, and the *e-h* excitation spectra are obtained by solving the Bethe-Salpeter equation (BSE) [28], which is transformed into a simplified eigenvalue problem after decoupling the excitations and de-excitations (also known as the Tamm-Dancoff approximation) [28]:

$$(E_{c\bar{k}} - E_{v\bar{k}})A^S_{vc\bar{k}} + \sum_{v'c'\bar{k}'} <vc\bar{k}|K^{eh}|v'c'\bar{k}'> A^S_{v'c'\bar{k}'} = \Omega^S A^S_{vc\bar{k}}, \quad (1)$$

where $K^{eh}$ is the *e-h* interaction kernel, $E_{c\bar{k}}$ ($E_{v\bar{k}}$) are the quasiparticle energy of conduction (valence) states, *S* labels the excitonic state $|S>$ with eigenvalue $\Omega^S$ and eigenstate function constructed using the electron and hole wave functions with the *e-h* coefficient $A_{vc\vec{k}}^{S}$:

$$\psi^{S}(\vec{r}_e,\vec{r}_h) = \sum_{vc\vec{k}} A_{vc\vec{k}}^{S} \varphi_{c\vec{k}}(\vec{r}_e)\varphi_{v\vec{k}}(\vec{r}_h). \tag{2}$$

In our calculations, four valence and four conduction states are included in the expansion of the excitonic wave functions, which should be adequate to cover excitations of up to at least 6 eV. We have also carried calculations using more valence and conduction bands; the results for low energy excitations are practically unchanged as shown in Supplemental Material (Table S3). The contribution to an excitonic state $|S>$ from *e-h* pairs in the BZ can be conveniently visualized using the *k*-dependent excitonic amplitude function $|A_{\vec{k}}^{S}|^2 = \sum_{vc}|A_{vc\vec{k}}^{S}|^2$. The imaginary part of the frequency-dependent macroscopic dielectric function including the excitonic effects is then given by (using atomic units)

$$\varepsilon_2(\omega) = \frac{16\pi^2}{\omega^2}\sum_{S}|\vec{\lambda}\cdot<0|\vec{v}|S>|^2\,\delta(\omega-\Omega^S), \tag{3}$$

where $\vec{\lambda}$ is the unit vector of the polarization direction of light and $\vec{v}$ is the velocity operator. A Gaussian smearing of width 25 meV is used in the calculation of $\varepsilon_2(\omega)$.

Figure 5 compares the calculated $\varepsilon_2(\omega)$ for monolayer (a), bilayer (b), and bulk (c) MoSi$_2$N$_4$. We include results calculated with (blue) and without (black) *e-h* interaction. Including the *e-h* interaction not only significantly shifts the absorption edge but also produces prominent excitonic absorption peaks. Note that it is well-known the dielectric function is volumetric, therefore it scales with the thickness of the vacuum layer for 2D systems[29-31]. However, the excitonic features (i.e., the position and relative strength of the excitonic peaks) can be compared directly with experiment as long as the results are adequately converged with respect to the interlayer separation. The first absorption peak (i.e., peak A) at 2.18 eV for the monolayer MoSi$_2$N$_4$ arises from two degenerate excitonic states. This excitation energy agrees well with the measured first optical absorption peak at 2.21 eV [13]. It is interesting that two excitonic states can give rise to such a strong absorption. The *k*-resolved *e-h* pair amplitudes $|A_{\vec{k}}^{S}|^2$ for these two excitonic states are shown in Fig. 5 (d). It is clear that the dominant contribution to the first two excitonic states comes from the *e-h* pairs near the minimum direct gap at the K and K' points [see Fig. 4 (a) for the quasiparticle band structure of monolayer MoSi$_2$N$_4$]. This is very similar to that in monolayer MoS$_2$ [32]. Therefore, for the A excitons, we can define a non-interacting electron-hole pair excitation gap $E^{\text{non-int}} = E^{QP}_{cK} - E^{QP}_{vK}$, which is the minimum quasiparticle band gap at the K point. Measuring from this non-interacting electron-hole pair excitation gap, we obtain the exciton binding energy of about 0.95 eV for the first two excitonic states.

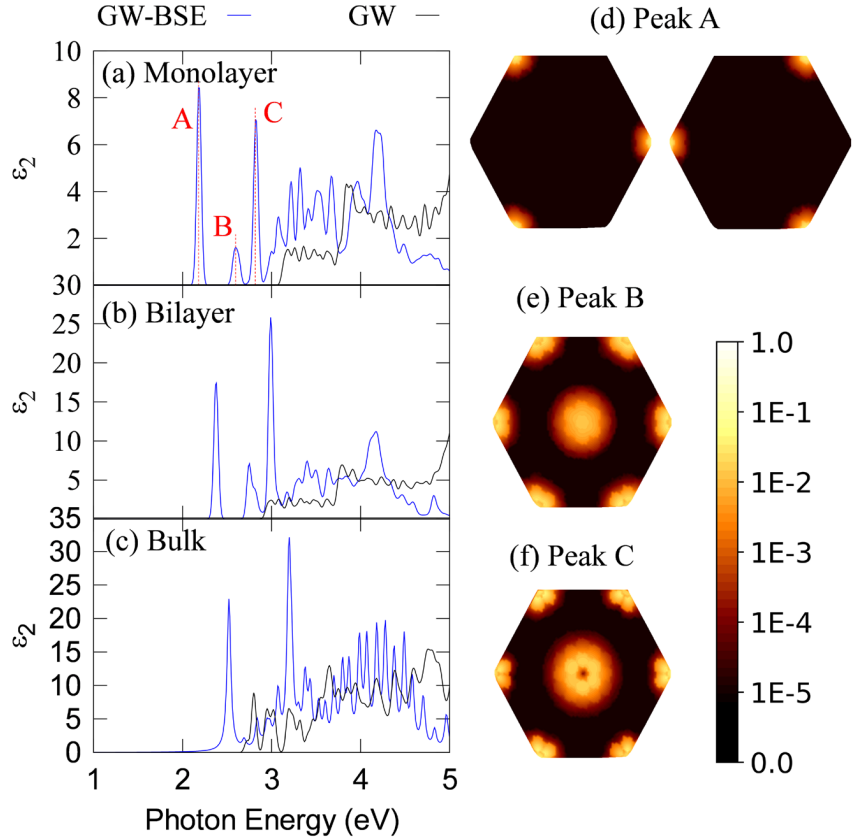

**Figure 5.** Layer-dependent absorption spectra (a-c) of MoSi$_2$N$_4$ and *e-h* pair amplitudes of the first three absorption peaks of monolayer MoSi$_2$N$_4$.

Peak B (at around 2.58 eV) consists of eight nearly degenerate excitonic states; these excitonic states are derived from *e-h* pairs around the Γ point and near the K (K') point with close quasiparticle band gaps. This can be clearly seen in Fig. 5 (e), which shows $\sum_{S \in \{B\}} |A^S_{\vec{k}}|^2$, summation of the *k*-resolved *e-h* amplitudes of the eight excitonic states that contribute to the B absorption peak. Another pronounced adsorption peak (peak C) appears at about 2.82 eV. This absorption peak comprises ten bright excitonic states. The summation of the *e-h* amplitudes, $\sum_{S \in \{C\}} |A_k^S|^2$, of these states is shown in Fig. 5 (f), suggesting that this absorption peak primarily originates from *e-h* pairs states with wave vectors near the Γ point along the Γ-M directions, forming a ring-like structure around the Γ point with small mixture of *e-h* pairs near the K (K') point. These features are consistent with the GW band structure shown in Fig. 4.

Figure 5 (b) and (c) show the imaginary part of the dielectric function of the bilayer and bulk phases. Similar to the case of monolayer, three strong low energy absorption peaks can be clearly identified for the bilayer system. Due to the increasingly stronger dielectric screening, therefore weakened *e-h* interaction, the first adsorption peaks of the bilayer and bulk phases actually move to higher energies (2.38 eV for bilayer and 2.52 eV for bulk), giving rise to exciton binding energies of 0.56 eV (bilayer) and 0.19 eV (bulk), to be compared with 0.95 eV for the monolayer. The blue shift of the optical gap for the bilayer and bulk structures (compared with the monolayer) is in stark contrast to other layered materials [5, 33], for which the optical gap typically decreases with increasing number of layers. This seemingly abnormal result can be traced to the fact that the band edge states of $MoSi_2N_4$ are largely protected from interlayer coupling as we have discussed in sections II A and B. As a result, the minimum (indirect) DFT band gap does not experience a significant reduction with increasing number of layers as shown in Table III. The calculated DFT direct gap at the K point actually increases slightly from 2.03 eV (monolayer) to 2.10 eV (bulk).

Finally, we summarize in Fig. 6 the evolution of indirect minimum band gap, the direct band gap at the K point, and the optical gap as a function of number of layers. As we have discussed earlier, the band gap of $MoSi_2N_4$ calculated at the PBE level is surprisingly stable thanks to the protected band edge states which are not susceptible to interlayer chemical coupling. Including the nonlocal self-energy correction within the GW approximation, however, leads to a significant layer-dependent quasiparticle band gap. Therefore, $MoSi_2N_4$ provides an interesting material system with robust band edge states that are spared from undesirable chemical coupling, but at the same time, tunable quasiparticle and optical properties through dielectric (or Coulomb) engineering.

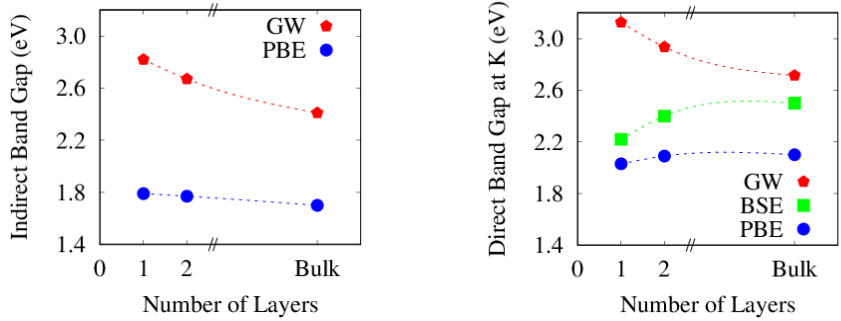

**Figure 6.** Evolution of the indirect minimum band gap (left panel), the direct band gap at the K point and the optical gap (right panel) as a function of the number of layers calculated at different theory levels.

Before we conclude, we would like to mention that calculations of *e-h* excitations at the GW-BSE level are extremely difficult to converge [23, 32, 34], especially with respect to the *k*-point sampling density. To alleviate the computational burden of BSE calculations, one often carries out the calculations on a relative coarse *k*-grid, the *e-h* kernel matrices are then interpolated to finer *k*-grids to obtain the final results. For the monolayer and bilayer structures, the *e-h* interaction kernel matrices are first calculated using a coarse 12×12×1 *k*-grid, the results are then interpolated onto a fine 72×72×1 *k*-grid from which the *e-h* excitations and optical absorption are obtained. For the bulk system, we use a 9×9×2 coarse *k*-grid and a 36×36×8 fine *k*-grid in our calculations. We have carefully tested the convergence of our results with respect to the *k*-point sampling density (Table S2 and Fig. S3 in Supplementary Material). We have also tested the adequacy of the number of valence and conduction bands included in the expansion of the excitonic wave functions (Table S3 in Supplementary Material).

## III. Conclusion

We have performed detailed first-principles calculations for the newly discovered layered material $MoSi_2N_4$. We find that the interlayer binding energy between $MoSi_2N_4$ layers is comparable to those in $MoS_2$ and other well-studied layered materials, but the calculated DFT-PBE band gap of $MoSi_2N_4$ is barely affected by the interlayer interaction, in stark contrast to other layered materials. Detailed analyses reveal that the wave functions of the band edge states of $MoSi_2N_4$ are derived primarily from interior atomic orbitals (i.e., Mo and interior N atoms) and thus are shield from surface or interfacial chemical perturbations. The quasiparticle and optical properties of $MoSi_2N_4$ can still be modulated through the nonlocal dielectric screening effects, as demonstrated by our GW-BSE calculations. The quasiparticle band gap of $MoSi_2N_4$ varies from 2.82 eV for the monolayer to 2.41 eV for the bulk phase. Interestingly, the optical band gap actually increases slightly with the increasing number of layers, changing from 2.19 eV (monolayer) to 2.52 eV (bulk). Our findings suggest that $MoSi_2N_4$ offers robust band edge states that are largely protected from undesirable environmental chemical interactions. The quasiparticle and optical properties can then be solely tuned by nonlocal screening effects, which can be achieved through the choice of appropriate substrate and/or controlling the layer thickness. This unique property, together with the moderate band gap and the thermodynamic and mechanical stability, may pave the way for applications of $MoSi_2N_4$ in areas including energy, 2D electronics, and optoelectronics.

## IV. Computational Methods

Structural optimizations for the monolayer, bilayer, and bulk phase $MoSi_2N_4$ with different layer stacking patterns are carried using the Vienna *ab initio* simulation package (VASP) [35, 36]. The Perdew-Burke-Ernzerhof (PBE) [37, 38] functional is used for the basic electronic band structure calculations, and the DFT-D3 correction of Grimme et al. [17] is employed in structural optimizations to account for the van der Waals (vdW) interactions. To minimize the fictitious interaction between periodic image layers, a vacuum layer of over 20 Å is included in the unit cells of monolayer and bilayer systems.

The layer-dependent quasiparticle and optical properties of $MoSi_2N_4$ are calculated using the BerkeleyGW package [39] within the GW [40] and Bethe-Salpeter equation (BSE) [28] approach. Norm-conversing pseudopotentials [41] are used for GW calculations, with semicore states (i.e., 4*s* and 4*p*) of Mo included as valence electrons. Cutoff for the plane wave expansion of the Kohn-Sham wave functions is set at 125 Ry, and that for the dielectric function is 50 Ry. We use the Hybertsen–Louie generalized plasmon-pole model (HL-GPP) [40] to extend the static dielectric function to finite frequencies. For GW calculations of monolayer and bilayer systems, the truncated Coulomb potential [42] is used to eliminate the artificial image interactions. We employed two recently implemented accelerated methods [25, 26] which greatly improve the efficiency of fully converged GW calculations for 2D systems. These two methods address two fundamental bottlenecks of GW calculations of 2D materials, leading to an overall speed-up factor of over three orders of magnitude compared with the conventional approaches as we have discussed in earlier publications [25, 26]. Other computational details have been discussed in Results and Discussion, and also in Supplemental Material.

**Acknowledgments**

This work is supported in part by the National Natural Science Foundation of China (Nos. 51632005, 51572167, and 11929401), the National Key Research and Development Program of China (No. 2017YFB0701600), Guangdong Innovative and Entrepreneurial Research Team Program (Grant No. 2019ZT08C044), and Shenzhen Science and Technology Program (KQTD20190929173815000). Work at UB is supported by the US National Science Foundation under Grant No. DMREF-1626967. W. Z. also acknowledges the support from the Guangdong Innovation Research Team Project (Grant No. 2017ZT07C062), and the Shenzhen Pengcheng-Scholarship Program. W. G. acknowledges the supports by the Fundamental Research Funds for the Central Universities, grant DUT21RC(3)033. We acknowledge the computational support provided by the Center for Computational Science and Engineering at Southern University of Science and Technology, the Center for Computational Research at UB, and Shanghai Supercomputer Center.

# Supplementary Material

## MoSi$_2$N$_4$: An emerging 2D electronic material with protected band edge states and dielectric tunable quasiparticle and optical properties

Yabei Wu,[1,2,3,‡] Zhao Tang,[4,‡] Weiyi Xia,[4] Weiwei Gao,[5] Fanhao Jia,[4,6] Yubo Zhang,[1,2] Wenguang Zhu,[3] Wenqing Zhang,[1,2,†] and Peihong Zhang[4,†]

1. Department of Physics and Shenzhen Institute for Quantum Science & engineering, Southern University of Science and Technology, Shenzhen, Guangdong 518055, China

2. Guangdong Provincial Key Lab for Computational Science and Materials Design, and Shenzhen Municipal Key Lab for Advanced Quantum Materials and Devices, Southern University of Science and Technology, Shenzhen, Guangdong 518055, China

3. ICQD, Hefei National Laboratory for Physical Science at the Microscale, Key Laboratory of Strongly-Coupled Quantum Matter Physics, Chinese Academy of Sciences, Department of Physics, and Synergetic Innovation Center of Quantum Information and Quantum Physics, University of Science and Technology of China, Hefei, Anhui 230026, China

4. Department of Physics, University at Buffalo, State University of New York, Buffalo, New York 14260, USA

5. Department of Physics, Dalian University of Technology, Dalian, Liaoning 116024, China

6. International Centre for Quantum and Molecular Structures, Materials Genome Institute, Department of Physics, Shanghai University, 99 Shangda Road, Shanghai 200444 China

[†]zhangwq@sustech.edu.cn; pzhang3@buffalo.edu

[‡]These authors contributed equally to this work.

## 1. Structural parameters of bilayer and bulk MoSi$_2$N$_4$ with different stackings

Figure S1 shows the six different stacking patterns of bilayer and bulk MoSi$_2$N$_4$ studied in this work. Table S1 shows the corresponding structural parameters. Configuration D is the most stable stacking pattern for both bilayer and bulk phases. The DFT-PBE band structures of the five metastable structures for the bilayer and bulk phases are shown in Fig. S2.

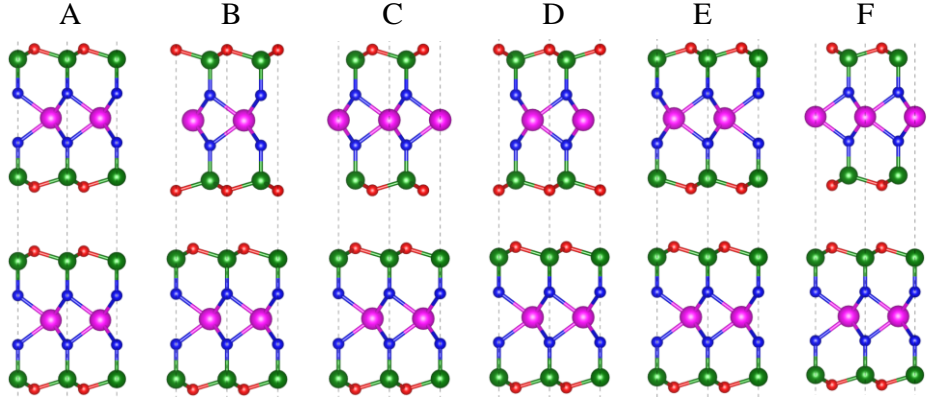

**Figure. S1.** Crystal structures of the six proposed bilayer/bulk configurations of MoSi$_2$N$_4$.

**Table S1.** Optimized lattice parameters (lattice constant *a* and interlayer distance *d*) for six bilayer and bulk MSi$_2$N$_4$ structures using the PBE functional with the DFT-D3 correction. The relative energies to the most energetic favorable structure (D) are also listed.

|   | Bilayer | | | Bulk | | |
|---|---|---|---|---|---|---|
|   | *a* (Å) | *d* (Å) | ΔE (meV/atom) | *a* (Å) | *d* (Å) | ΔE (meV/atom) |
| A | 2.894 | 3.270 | 7.4 | 2.893 | 3.222 | 14.9 |
| B | 2.895 | 2.882 | 1.1 | 2.894 | 2.861 | 2.3 |
| C | 2.895 | 2.882 | 1.3 | 2.894 | 2.866 | 2.3 |
| **D** | **2.896** | **2.825** | **0.0** | **2.896** | **2.807** | **0.0** |
| E | 2.894 | 2.911 | 2.8 | 2.893 | 2.911 | 5.6 |
| F | 2.895 | 3.193 | 7.2 | 2.893 | 3.195 | 5.6 |

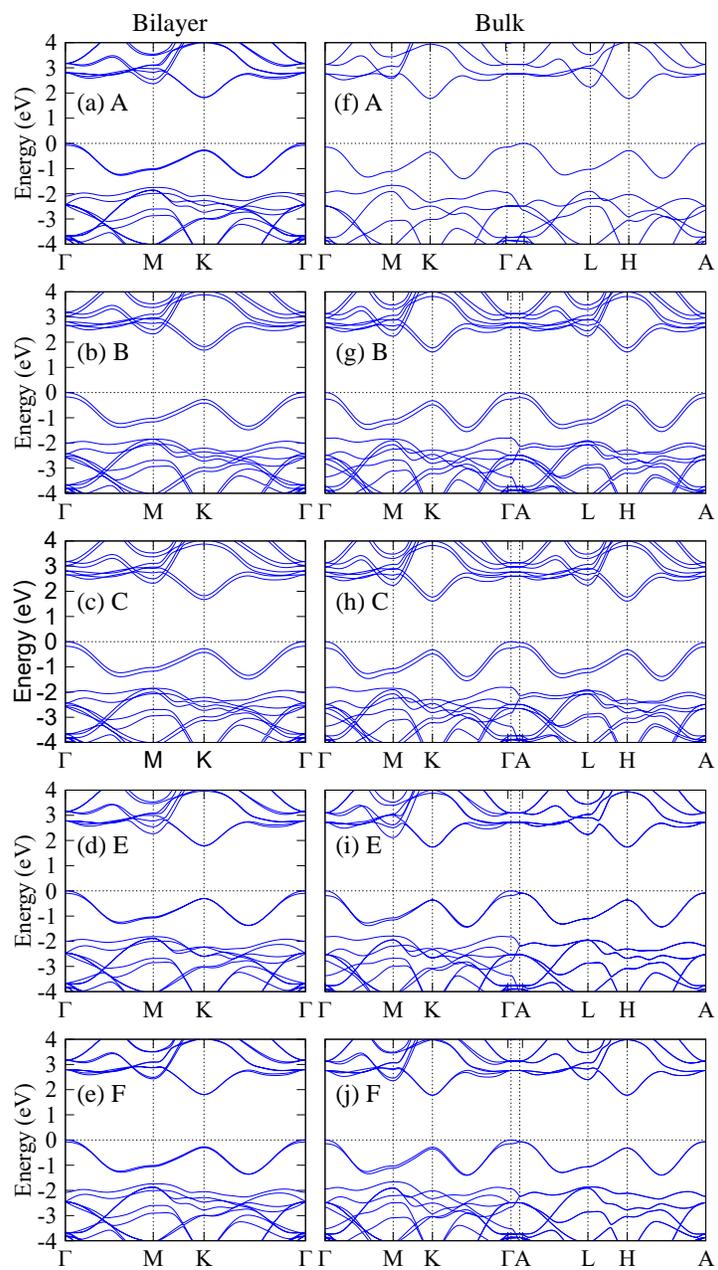

**Figure. S2.** DFT-PBE band structures of the other five bilayer (left panel) and five bulk (right panel) configurations.

## 2. Convergence tests of GW-BSE results

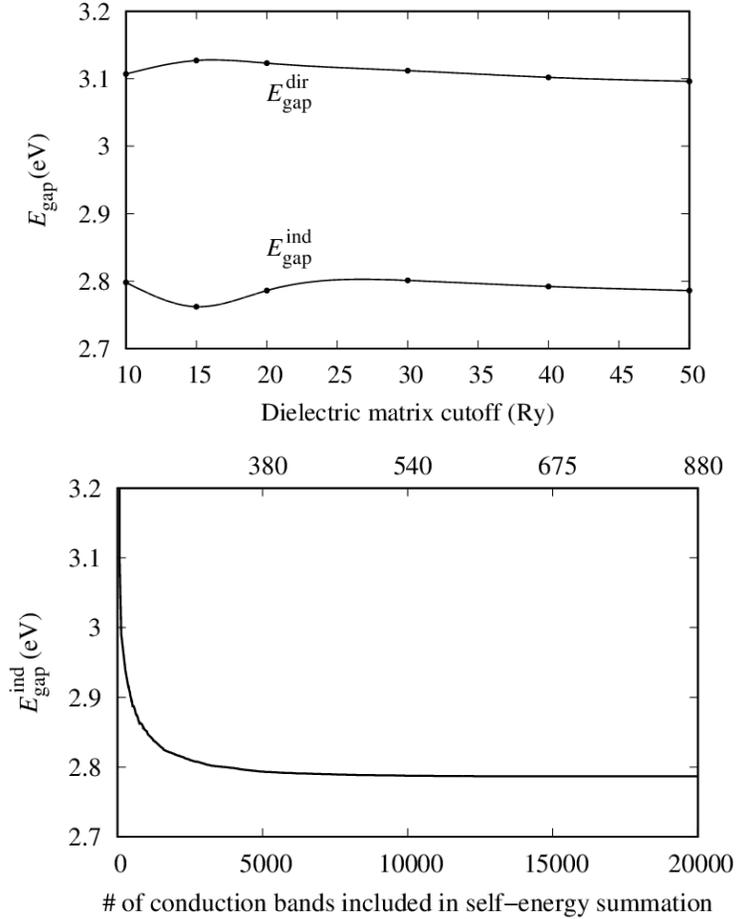

**Figure S3.** Convergence behavior of the calculated GW band gap for monolayer MoSi$_2$N$_4$ with respect to the kinetic energy cutoff for the dielectric matrices (upper) and the number of conduction bands included in the GW calculation (bottom).

Figure S3 shows the convergence behavior of the calculated GW band gap of monolayer MoSi$_2$N$_4$ with respect to the kinetic energy cutoff for the dielectric matrices (the upper panel) and the number of empty states (the bottom panel) included in the GW calculations. The lower horizontal axis of the bottom panel shows the number of bands effectively included in our calculation, whereas the upper horizontal axis shows the number of integration points for evaluating the band summation using our new energy integration method [1], suggesting a speed-up factor of more than 20. The final results we reported in the manuscript are calculated using a 50 Ry cutoff for the dielectric matrix and effectively including *all* (more than 30,000) conduction bands, which should give well-converged results with respect to these two parameters.

Fully converged GW calculations for 2D materials are more challenging than those for their 3D counterparts. GW (and BSE) results should also be checked against other convergence parameters such as *k*-point density for the Brillouin zone integration and the thickness of the vacuum layer included in the calculation (to reduce the spurious interlayer coupling). Both issues have been well-recognized and discussed in literature. To reduce the dependence of the calculated results on the thickness of the vacuum layer, we use the Coulomb truncation scheme proposed by Ismail-Beigi [2]. However, even with this truncation scheme, one still needs to include a sizeable vacuum layer in the calculation. In our original manuscript, the $c$ lattice constant (perpendicular to the layer direction) for the GW/BSE calculations of monolayer $MoSi_2N_4$ is set at 30 Å, which, from our experience, is large enough to converge the results to within 0.1 eV. We then use our newly developed mini-BZ subsampling and analytical integration method [3] to evaluate the BZ integration of the self-energy.

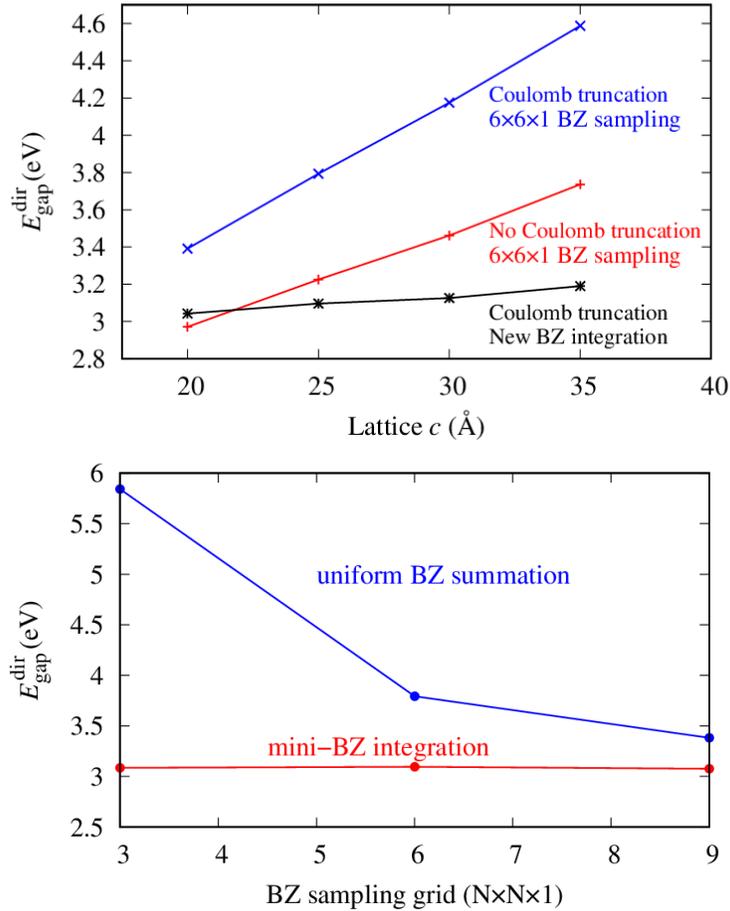

**Figure S4.** Convergence behavior of the calculated GW band gap for monolayer $MoSi_2N_4$ with respect to the c lattice constant (upper panel) and the BZ integration technique (bottom panel).

The upper panel of Fig. S4 shows the calculated direct band gap of monolayer $MoSi_2N_4$ as a function of lattice $c$ using different approaches. The blue curve shows the results calculated with a truncated Coulomb potential using a 6×6×1 uniform $k$-point sampling, whereas the red curve shows the result calculated using the 1/r long-range Coulomb potential with 6×6×1 uniform $k$-grid. Interestingly, using a truncated Coulomb potential would seem to give worse results if the BZ integration is not converged with respect to the *k*-point sampling density (i.e., using a uniform 6×6×1 grid). In other words, with the use of a truncated Coulomb potential, the GW results converge *slower* with respect to the BZ *k*-point sampling density than using a long-range Coulomb potential. Obviously, neither results are converged. This subtle interplay among the convergence of the BZ integration, the use (or not) of truncated Coulomb potential, and the thickness of the vacuum layer has been discussed in literature, and our results are consistent with the results of Rasmussen et al. [4]. The black curve in the upper panel shows GW results as a function of the *c* lattice constant using our new BZ integration technique [3]. We can confidently say that the results converge to about 0.1 eV with respect to the *c*-lattice. Finally, the bottom panel shows the calculated GW band gap as a function of BZ *k*-point sampling density, with and without the use of the mini-BZ integration technique we developed [3], using a lattice constant *c* of 30 Å. Using our newly developed method, the calculated band gap converges even with a very coarse k-grid 3×3×1. In comparison, conventional BZ integration using a uniform *k*-grid would likely require an 18×18×1 (or denser) grid to achieve the same level of convergence. The results reported in our original manuscript were calculated using a 6×6×1 *k*-grid (with our new mini-BZ integration) and a *c* lattice constant of 30 Å.

Table S2 shows the positions and exciton binding energies of the first excitonic peaks of monolayer, bilayer, and bulk $MoSi_2N_4$ with respect to the *k*-point density. The final results that we present in the main text should converge to within 50 meV with respect to the *k*-grid density for the monolayer and bilayer systems. For the bulk phase, the results should convergence to within 10 meV. Figure S5 shows the real and imaginary parts of the dielectric function for monolayer $MoSi_2N_4$ based on GW-BSE calculations using different fine *k*-grids. Finally, Table S3 shows the variation of the calculated positions of the first three excitonic peaks of monolayer $MoSi_2N_4$ with different number of valence and conduction bands included in the BSE calculations.

**Table S2.** Convergence behavior of the positions and exciton binding energies of the first excitonic peaks of monolayer, bilayer, and bulk $MoSi_2N_4$ with respect to the *k*-point density.

| System | Coarse grid | Fine grid | First Peak Position (eV) | $E_b$ (eV) |
|---|---|---|---|---|
| Monolayer | 12×12×1 | 24×24×1 | 2.28 | 0.85 |
|  |  | 36×36×1 | 2.22 | 0.91 |
|  |  | 72×72×1 | 2.18 | 0.95 |
| Bilayer | 12×12×1 | 24×24×1 | 2.47 | 0.46 |
|  |  | 36×36×1 | 2.40 | 0.53 |
|  |  | 72×72×1 | 2.37 | 0.56 |
| Bulk | 9×9×2 | 12×12×3 | 2.48 | 0.23 |
|  |  | 18×18×4 | 2.50 | 0.21 |
|  |  | 24×24×6 | 2.52 | 0.19 |
|  |  | 36×36×8 | 2.52 | 0.19 |

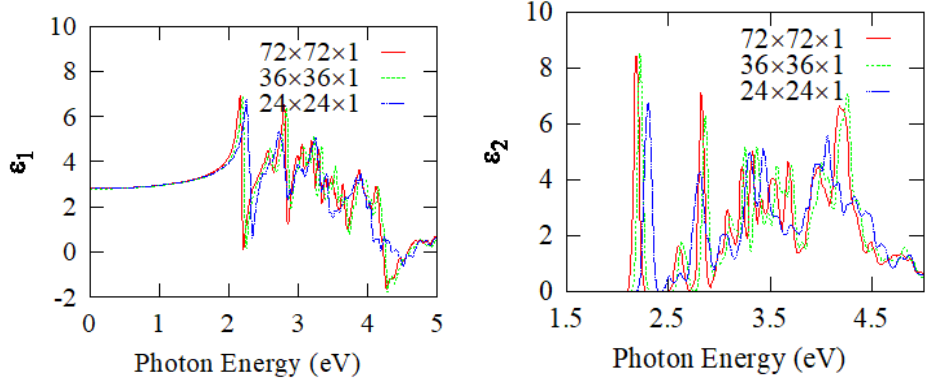

**Figure. S5.** The real and imaginary parts of the dielectric function for monolayer MoSi$_2$N$_4$ based on GW-BSE calculations using different fine *k*-grids.

**Table S3.** Variation of the calculated positions of the first three excitonic peaks of monolayer MoSi$_2$N$_4$ with different number of valence and conduction bands included in the BSE calculations.

| # of valence bands | # of conduction bands | Peak A (eV) | Peak B (eV) | Peak C (eV) |
|---|---|---|---|---|
| 4 | 4 | 2.18 | 2.58 | 2.82 |
| 11 | 6 | 2.18 | 2.59 | 2.82 |